\documentclass[twocolumn,floatfix, showpacs, nofootinbib]{revtex4}

\usepackage[final]{epsfig}
\usepackage{amsmath}
\usepackage{parskip}
 
\def\lsim{\,\raise0.3ex\hbox{$<$\kern-0.75em\raise-1.1ex\hbox{$\sim$}}\,}
\def\gsim{\,\raise0.3ex\hbox{$>$\kern-0.75em\raise-1.1ex\hbox{$\sim$}}\,}
 
\begin{document}
 
\title{The form of the elastic energy loss probability distribution in a static medium}
 
\author{Jussi Auvinen}
\email{jussi.a.m.auvinen@student.jyu.fi}
\author{Thorsten Renk}
\email{trenk@phys.jyu.fi}
\affiliation{Department of Physics, P.O. Box 35, FI-40014 University of Jyv\"askyl\"a, Finland}
\affiliation{Helsinki Institute of Physics, P.O. Box 64, FI-00014, University of Helsinki, Finland}
 
\pacs{25.75.-q, 25.75.Bh}

\begin{abstract}
We examine the probability distributions $P(E,t)$ of the energy of a hard parton traveling in a partonic medium of constant density for a time $t$ while undergoing elastic 2 $\rightarrow$ 2 pQCD interactions using a Monte-Carlo model. The form of these distributions is found to be non-Gaussian, confirming results by other groups with similarly detailed models and challenging the validity of the widely used diffusion approximation in elastic energy loss modeling.
\end{abstract}

\maketitle

\section{Introduction}

One of the key observables in ultrarelativistic heavy ion experiments is the strong suppression of hadrons with high transverse momentum $P_T$ in heavy-ion (A-A) collisions as compared with a p-p baseline. This has now been measured both at RHIC \cite{Adler:2006hu} and the LHC \cite{Aamodt:2010jd}. This suppression is typically expressed in terms of the nuclear modification factor 

\begin{equation}
R_{AA}(P_T,y,{\bf b}) = \frac{dN^{\pi}_{AA}/dP_Tdy}{T_{AA}({\bf b}) d\sigma^{pp}/dP_Tdy}.
\end{equation}

The transverse momentum dependence of $R_{AA}$ is notably more pronounced at the LHC than at RHIC, which is expected on rather general grounds due to the harder partonic $p_T$ spectrum \cite{Renk:2007jv} which can be understood as a filter through which the energy loss probability distribution $P(\Delta E)$ is observed. A more pronounced $P_T$ dependence of $R_{AA}$ in turn implies a higher sensitivity to details of $P(\Delta E)$. Because of this increased sensitivity to the energy loss probability distribution at the LHC, it is crucial for a model not only to reproduce the correct mean energy loss but also the correct fluctuations around the mean.

The physics mechanism of the energy loss may be medium-induced radiation or elastic collisions with medium constituents. A sizeable elastic contribution to the total energy loss has been invoked by various groups \cite{Mustafa:2003vh,DuttMazumder:2004xk,Roy:2005am,Wicks:2005gt,Djordjevic:2006tw,Majumder:2008zg,Zapp:2008gi}. In such calculations, fluctuations of the elastic energy loss often are treated in a diffusion (Gaussian) approximation. In this approximation, a purely elastic picture reproduces the $P_T$-dependence of $R_{AA}$ quite well and is in agreement with the LHC data \cite{Renk:2011gj,Renk:2010mf}. However, this agreement is misleading, as in a calculation relaxing the diffusion assumption the form of $R_{AA}(P_T)$ resembles more the results of the radiative energy loss calculations and hence such a computation without diffusion approximation is in disagreement with the data \cite{Auvinen:2011wx}. In this paper, we illustrate that a large difference between the treatment in a phenomenological and a detailed Monte-Carlo (MC) model is the strongly non-Gaussian shape of $P(\Delta E)$.

We carry out our study using a static QCD medium of constant temperature (and density). While unrealistically simple, it nevertheless provides an useful setting for exploring the evolution of energy loss probability distributions, as well as providing a common ground on which to compare the results of different models. We first examine how the shape of parton energy distribution depends on the assumptions made about the similarity of successive scatterings. We then compare the results of our model to those of Boltzmann Approach to MultiParton Scatterings (BAMPS) \cite{Fochler:2010wn} and results by Schenke {\it et al.} \cite{Schenke:2009ik}.

\section{The Monte Carlo simulation}

Our energy loss model is based on the 2-to-2 scattering rates for a high-energy parton of a type $i$ \cite{Auvinen:2009qm}, 

\begin{equation}
\label{scattrate}
\Gamma_{ij\rightarrow kl} = \frac{1}{16\pi^2E_1^2}\int_{\frac{m^2}{2E_1}}^{\infty}dE_2f_j(E_2,T) \Omega_{ij\rightarrow kl}(E_1,E_2,m^2),
\end{equation}
where
\begin{equation}
\label{omegafunction}
\Omega_{ij\rightarrow kl}(E_1,E_2,m^2)=\int_{2m^2}^{4E_1E_2}d\hat{s} [\hat{s}\sigma_{ij\rightarrow kl}(\hat{s})].
\end{equation}

The energy of the high-energy parton $i$ is $E_1$ and $E_2$ is the energy of the thermal quark (gluon) $j$ with a Fermi-Dirac (Bose-Einstein) distribution function $f_j(E_2,T)$. The scattering cross section $\sigma_{ij\rightarrow kl}(\hat{s})$ depends on the standard Mandelstam variable $\hat{s}$. A regulator $m=s_m\sqrt{4\pi \alpha_s}T$ is introduced for the infrared singularities appearing in the cross section. The strong coupling constant $\alpha_s$ is kept fixed while $s_m$ is a parameter typically of the order of one. 

The propagation of the hard parton through the plasma is done in small time steps $\Delta t$. The Poisson probability for not colliding in this time interval is
\begin{equation}
P(\text{No collisions in } \Delta t)=e^{-\Gamma_i \Delta t}, 
\end{equation}

where $\Gamma_i$ is the sum of all possible scattering rates for the hard parton of the type $i$. By keeping $\Delta t$ small, we can safely assume that there will be at most one collision on this time interval. What type of scattering process happens is sampled according the ratios of the partial scattering rates \eqref{scattrate} to the total scattering rate. 

In order to illustrate the effect of a diffusion approximation, we use in this study two different methods for treating the momentum exchange between the hard parton and the medium particle. The standard one is to sample the Mandelstam-$\hat{t}$ from the $\frac{d\sigma}{d\hat{t}}$ -distribution. The other method used here corresponds to the diffusion approximation and requires us to calculate the average Mandelstam-$\hat{t}$,

\begin{equation}
\langle \hat{t} \rangle = \frac{\int_{-\hat{s}+m^2}^{-m^2}d\hat{t} \, \hat{t} \frac{d\sigma}{d\hat{t}}}{\int_{-\hat{s}+m^2}^{-m^2}d\hat{t} \frac{d\sigma}{d\hat{t}}}. 
\end{equation}

It is then assumed that all scattering processes happen with this average momentum exchange. Since $\langle \hat{t} \rangle$ defined this way is not a constant but depends on Mandelstam-$\hat{s}$, we additionally use the average values for the energy of a thermal parton $\langle E_2 \rangle=3T$ and the collision angle $\langle \cos\theta_{12} \rangle \approx -\frac{1}{3}$ taken from \cite{Auvinen:2009qm}. This leaves $\langle \hat{s} \rangle$ dependent only on $E_1$. In order to keep the approximation scheme simple, we also approximate the amplitude of the process $qg \rightarrow qg$ with a pure $\hat{t}$-channel: $|M|_{qg \rightarrow qg}^2 \propto \frac{\hat{s}^2+\hat{u}^2}{\hat{t}^2}$. This approximation will decrease $\langle \hat{t} \rangle$ slightly. This is of little consequence, as this study focuses on the variation in the form of the probability distributions and not in the magnitude of the average energy loss. With these approximations, $\langle \hat{t} \rangle$ varies $ \sim 15\%$ in the energy range $E_1 = 25 - 50$ GeV at $T=300$ MeV. 

In both schemes, after scattering the final state parton with highest energy is chosen as the new hard parton to be propagated further. The procedures outlined above are repeated for a predetermined time period $t$ (corresponding to a length $L$), after which the simulation ends. Repeating the simulation several times produces a probability distribution for the energy of the hard parton at the time $t$, $P(E,t)=\frac{1}{N_{tot}} \frac{dN(E,t)}{dE}$.

\section{Results}

\begin{figure*}[!t]
\centering
\includegraphics[width=18.0cm]{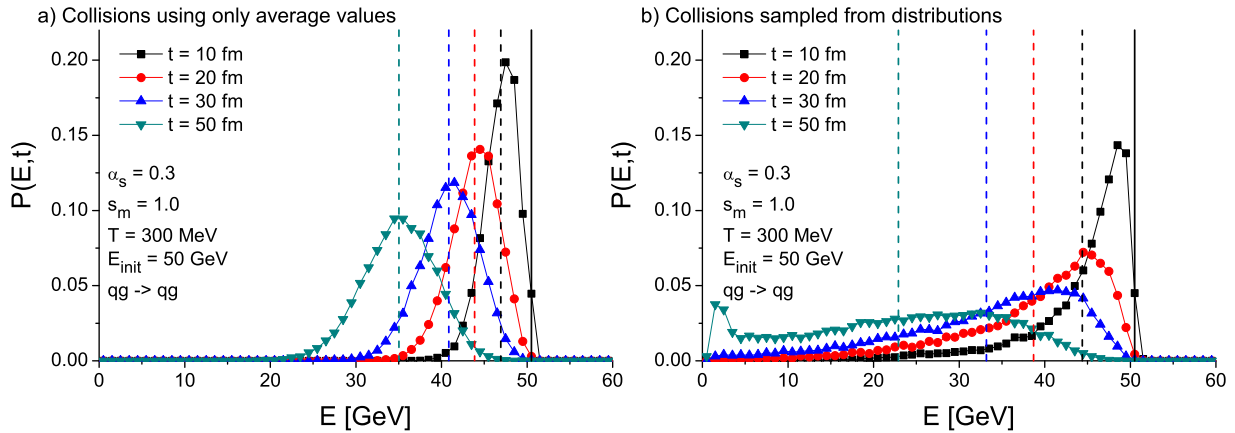}
\vspace{-0.2cm}
\caption{(Color online) a) Energy distribution of a 50 GeV quark after traveling in gluon plasma for various times, using average $\langle \hat{t} \rangle$ (see text). b) Energy distribution of a 50 GeV quark after traveling in gluon plasma for various times, values of the $\hat{t}$ sampled from the cross section distribution. Dashed vertical lines illustrate the average energy values, while the solid line marks the initial energy bin. The temperature of the medium is $T=300$ MeV, the strong coupling constant value is $\alpha_s=0.3$ and the regulator parameter $s_m$ is unity.}
\label{fig: T300_comparison}
\end{figure*}

\begin{figure}[!h]
\centering
\includegraphics[width=9.0cm]{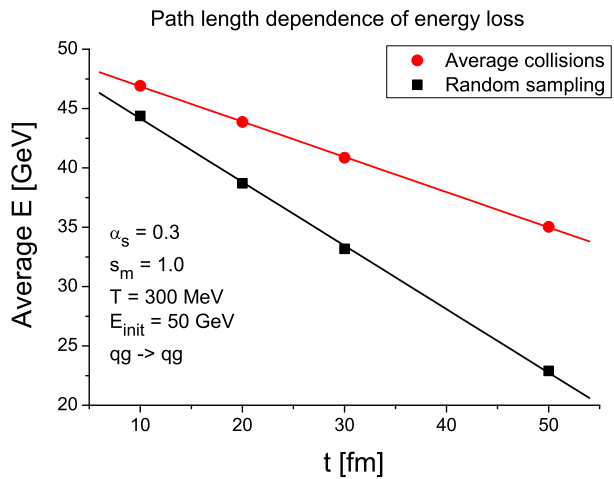}
\vspace{-0.2cm}
\caption{(Color online) Comparison of the hard parton average energy pathlength dependence in average collisions scheme and the distribution sampling scheme. Solid circles correspond with the dashed lines shown in Fig. \ref{fig: T300_comparison} a) and solid squares correspond with the dashed lines shown in Fig. \ref{fig: T300_comparison} b). Temperature of the medium is $T=300$ MeV, the strong coupling constant value is $\alpha_s=0.3$ and the regulator parameter $s_m$ is unity.}
\label{fig: T300_pathlength}
\end{figure}

We first examine the parton energy loss in a gluon plasma. Figure \ref{fig: T300_comparison} shows the difference between the two scenarios described in the previous section for an initial 50 GeV quark traveling in the medium for various times. Treating every collision as average event results in nearly Gaussian probability distributions for the parton energy, while in the scenario where we include the correct fluctuations by sampling the the thermal distributions and differential cross section the energy probabilities have a long tail towards lower energies. However, as we see in Figure \ref{fig: T300_pathlength}, the dependence of the average energy loss on the in-medium pathlength $L$ is linear in both cases. 

Next we select a gluon instead of a quark as initial parton and compare our results with the BAMPS \cite{Fochler:2010wn} for a $T=400$ MeV purely gluonic medium (i.e. we probe the channel process $gg \rightarrow gg$ only) and using the same coupling constant value $\alpha_s=0.3$. We find the qualitative behavior very similar, but quantitatively our model produces stronger energy loss. This difference can be largely attributed to the differences in the cross section regularization, which according to Zapp {\it et al.} can produce a factor two difference in the cross section \cite{Zapp:2008gi}.

\begin{figure}[!h]
\centering
\includegraphics[width=9.0cm]{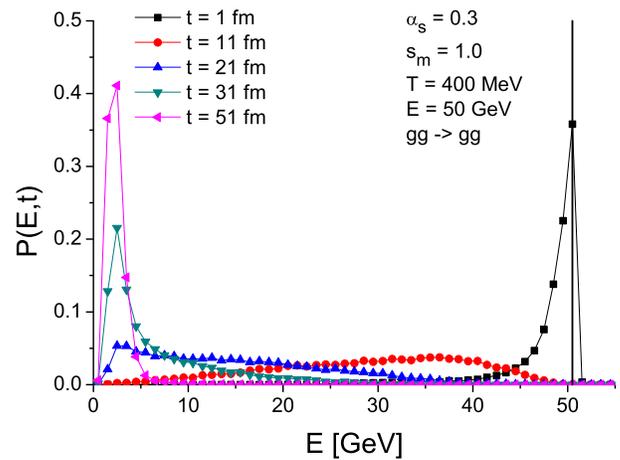}
\vspace{-0.2cm}
\caption{(Color online) Energy distribution of a 50 GeV gluon after various times, values of the $\hat{t}$ sampled from the cross section distribution. The vertical solid line marks the initial energy bin. The temperature of the gluonic medium is $T=400$ MeV, the strong coupling constant value is $\alpha_s=0.3$ and the regulator mass parameter is $s_m=1.0$. Compare with upper panel of Fig. 4 in Ref. \cite{Fochler:2010wn}.}
\label{fig: T400}
\end{figure}

Finally we study a 10 GeV quark traveling through quark-gluon plasma with temperature $T=200$ MeV, keeping the coupling strength $\alpha_s=0.3$, and compare our results with those by Schenke {\it et al.} \cite{Schenke:2009ik}. The results in this case agree both qualitatively and quantitatively. The main difference, i.e. the sudden drop in the 9.8-10.0 GeV bin, is once again explained by the chosen regulator as we, unlike Schenke {\it et al.}, ignore scattering events below the soft energy scale $\sim \sqrt{4 \pi \alpha_s}T$ and thus artificially suppress the number of particles with very small energy loss. Interestingly, our average energy loss is actually smaller than in Ref. \cite{Schenke:2009ik} in this scenario.

\begin{figure}[h]
\centering
\includegraphics[width=9.0cm]{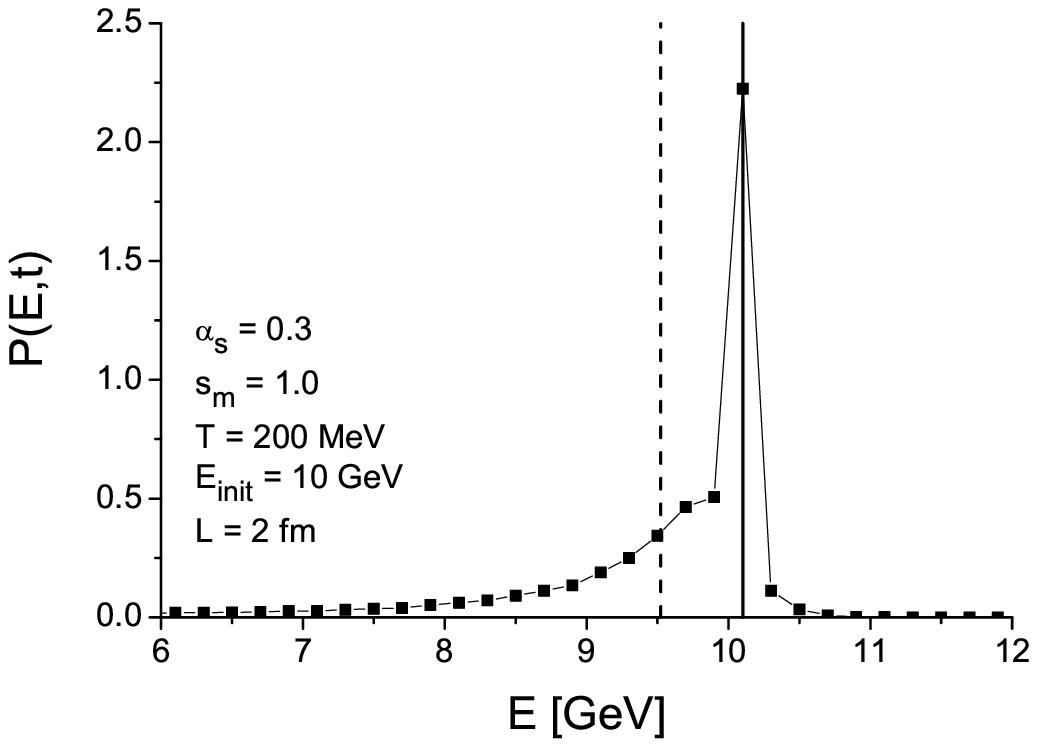}
\vspace{-0.2cm}
\caption{Energy distribution of a 10 GeV quark after traveling length $L=2$ fm in a quark-gluon medium with a temperature $T=200$ MeV, strong coupling constant value $\alpha_s=0.3$ and mass parameter value $s_m=1.0$. The dashed vertical line illustrates the average energy and the solid line the initial energy bin. Compare with Fig. 3 in Ref. \cite{Schenke:2009ik}.}
\label{fig: T200_L2}
\end{figure}

\section{Discussion}

We observe a strongly non-Gaussian shape of the energy (loss) probability distribution $P(\Delta E)$ of hard partons after elastic collisions with a QCD medium of constant density. The non-Gaussian nature is mainly driven by fluctuations around the average energy transfer per collision, other effects like parton type changing conversion reactions with the medium ($q\overline{q} \rightarrow gg$) which are often neglected in approximations to the elastic energy loss problem do not play a role.

Our results agree with observations previously made by Schenke {\it et al.} in Ref. \cite{Schenke:2009ik}. The emerging picture cast a serious doubt on the validity of the diffusion approximation in the elastic energy loss modeling. 

The detailed functional form of $P(\Delta E)$ is not an academic issue provided that the mean energy loss can be computed correctly, on the contrary it is directly linked to the $P_T$ dependence of $R_{AA}$ at LHC kinematics. The implication of our results is that a large elastic contribution to the total energy loss ($>10$\%) is not indicated by the LHC data. Initially results using the diffusion approximation gave a reasonable agreement with the observed $P_T$ dependence whereas radiative scenarios underpredicted the growth of $R_{AA}$ with $P_T$ \cite{Renk:2011gj}. This allowed for the possibility that a sufficiently large elastic contribution to the total energy loss could result in a scenario in agreement with the data. However, since a correct treatment of the fluctuations leads to the same underprediction of the rise of $R_{AA}$ with $P_T$ for elastic and radiative energy loss, this argument can no longer be made.

We also find that the pathlength dependence of elastic energy loss remains linear (as expected) in a constant medium, no matter how fluctuations are treated, thus all arguments made previously in the context of a diffusion approximation (e.g. \cite{ElasticPhenomenology}) based on pathlength to constrain the relative contribution of elastic energy loss to be about 10\% remain valid. Note also that dihadron correlations allow to constrain the magnitude of elastic energy loss \emph{from below} and are consistent with a 10\% contribution \cite{IAA_elastic}. There is thus growing evidence that the relative contribution of elastic energy loss is considerably smaller than expected by a straightforward computation based on an ideal quark-gluon gas picture.

\begin{acknowledgments}
J.A. gratefully acknowledges the grant from the Jenny and Antti Wihuri Foundation. T.R. is supported by the Academy Researcher program of the Finnish Academy (Project 130472) and from Academy Project 133005.
\end{acknowledgments}

\begin{figure}[!t]
\centering
\includegraphics[width=9.0cm]{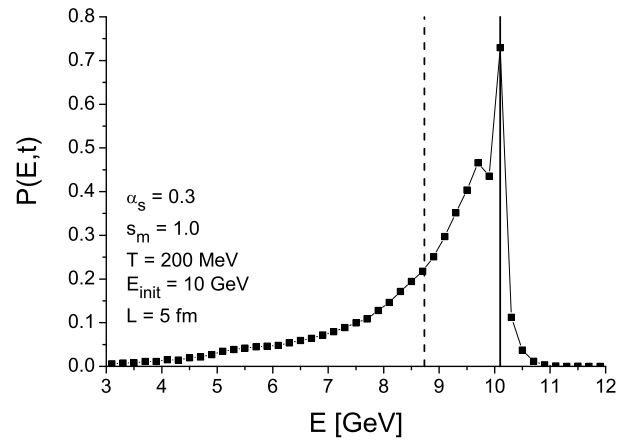}
\vspace{-0.2cm}
\caption{Energy distribution of a 10 GeV quark after traveling length $L=5$ fm in a quark-gluon medium with a temperature $T=200$ MeV, strong coupling constant value $\alpha_s=0.3$ and mass parameter value $s_m=1.0$. The dashed vertical line illustrates the average energy and the solid line the initial energy bin. Compare with Fig. 4 in Ref. \cite{Schenke:2009ik}.}
\label{fig: T200_L5}
\end{figure}

\pagebreak


\begin{thebibliography}{99}

\bibitem{Adler:2006hu}
  S.~S.~Adler {\it et al.}  [PHENIX Collaboration],
  Phys.\ Rev.\ Lett.\  {\bf 96}, 202301 (2006)

\bibitem{Aamodt:2010jd}
  K.~Aamodt {\it et al.}  [ALICE Collaboration],
  Phys.\ Lett.\  B {\bf 696}, 30 (2011).

\bibitem{Renk:2007jv}
  T.~Renk, K.~J.~Eskola,
  PoS {\bf LHC07}, 032 (2007).

\bibitem{Mustafa:2003vh}
  M.~G.~Mustafa, M.~H.~Thoma,
  Acta Phys.\ Hung.\  {\bf A22}, 93-102 (2005).

\bibitem{DuttMazumder:2004xk}
  A.~K.~Dutt-Mazumder, J.~-eAlam, P.~Roy, B.~Sinha,
  Phys.\ Rev.\  {\bf D71}, 094016 (2005).

\bibitem{Roy:2005am}
  P.~Roy, A.~K.~Dutt-Mazumder, J.~-eAlam,
  Phys.\ Rev.\  {\bf C73}, 044911 (2006).

\bibitem{Wicks:2005gt}
  S.~Wicks, W.~Horowitz, M.~Djordjevic, M.~Gyulassy,
  Nucl.\ Phys.\  {\bf A784}, 426-442 (2007).

\bibitem{Djordjevic:2006tw}
  M.~Djordjevic,
  Phys.\ Rev.\  {\bf C74}, 064907 (2006).

\bibitem{Majumder:2008zg}
  A.~Majumder,
  Phys.\ Rev.\  {\bf C80}, 031902 (2009).

\bibitem{Zapp:2008gi}
  K.~Zapp, G.~Ingelman, J.~Rathsman, J.~Stachel, U.~A.~Wiedemann,
  Eur.\ Phys.\ J.\  {\bf C60}, 617-632 (2009).

\bibitem{Renk:2011gj}
  T.~Renk, H.~Holopainen, R.~Paatelainen, K.~J.~Eskola,
  Phys.\ Rev.\  {\bf C84}, 014906 (2011).

\bibitem{Renk:2010mf}
  T.~Renk,
  Phys.\ Rev.\  {\bf C83}, 024908 (2011).

\bibitem{Auvinen:2011wx}
  J.~Auvinen, K.~J.~Eskola, H.~Holopainen, T.~Renk,
  J.\ Phys.\ G {\bf G38}, 124160 (2011).

\bibitem{Fochler:2010wn}
  O.~Fochler, Z.~Xu, C.~Greiner,
  Phys.\ Rev.\  {\bf C82}, 024907 (2010).

\bibitem{Schenke:2009ik}
  B.~Schenke, C.~Gale, G.~-Y.~Qin,
  Phys.\ Rev.\  {\bf C79}, 054908 (2009).

\bibitem{Auvinen:2009qm}
  J.~Auvinen, K.~J.~Eskola and T.~Renk,
  Phys.\ Rev.\  C {\bf 82}, 024906 (2010).
  
\bibitem{ElasticPhenomenology}
  T.~Renk,
  Phys.\ Rev.\ C\ {\bf 76} (2007) 064905.

\bibitem{IAA_elastic}
  T.~Renk,
  arXiv:1110.2313 [hep-ph].


\end{thebibliography}
\end{document}